\documentclass[12pt,eqsecnum]{JHEP3}
\title{Coherent States and N Dimensional Coordinate Noncommutativity}
\author{Myron Bander \\Department of Physics and Astronomy\\
University of California\\ Irvine, 
California 92697-4575, USA \\E-Mail: \email{mbander@uci.edu}}

\abstract{Considering coordinates as operators whose measured
values are expectations between generalized coherent states based on
the group $SO(N,1)$ leads to coordinate noncommutativity together with
full $N$ dimensional rotation invariance. Through the introduction of
a gauge potential this theory can additionally be made invariant under
$N$ dimensional translations. Fluctuations in coordinate measurements
are determined by two scales. For small distances these fluctuations
are fixed at the noncommutativity parameter while for larger distances
they are proportional to the distance itself divided by a {\em very}
large number. Limits on this number will lbe available from LIGO
measurements.}
\preprint{UCI-TR 2005-36}
\begin{document}

\section{Introduction}\label{intro} Most of the recent work on
coordinate noncommutativity has restricted itself to two spacial
dimensions, where the commutator $[x,y]=i\theta$ is implemented
through the Groenewold-Moyal \cite{GM} star product; $\theta$ is
constant for flat two-geometries and somewhat more complicated for
spherical ones \cite{Madore,MB1}. In dimensions higher than two this
procedure clearly breaks rotational invariance. One of the purposes of
this work is to introduce coordinate noncommutativity for unbounded
$N$ dimensional spaces and yet maintain $SO(N)$ invariance for the
resulting dynamics. In fact, through the introduction of a gauge
potential the dynamics can be made invariant invariant under $SO(N,1)$
transformations. The extra $N$ symmetries, those beyond $SO(N)$,
should not be viewed as usual Lorentz transformations as they do not
involve time, but rather as ``translations''. Implementation of such
translation symmetries will necessitate the introduction of a gauge
potential. The present treatment differs from ones in which Poincar\'e
symmetry is achieved through the introduction of a ``twist''
\cite{Chaichian:2004za,Wess:2003da} into the action of that algebra
while maintaining noncommutativity on a two dimensional plane. The
noncommutativity discussed here is fully $N$ dimensional and involves
an algebra larger than the one generated by the coordinates. Rotation
symmetry is achieved without deforming the angular momentum algebra.

As in ordinary quantum mechanics the coordinates are operators and
their measured values are determined by which states matrix elements are
taken in. For ordinary, coordinate commuting, quantum mechanics we may
choose these states to be simultaneous eigenstates of these
coordinates; noncommutativity precludes having a simultaneous
eigenstate of all the coordinate operators. Coherent states
\cite{Perelomov} are those that minimize momentum-position uncertainty
and it is tempting to use their analogs for the problem of coordinate
noncommutativity. Such states have been applied to investigations
of noncommutative geometries \cite{prevnc}, principally the ones induced
by the star product rules. 

One of the primary properties of coherent states is their minimization
of the uncertainty in the measurement of noncommuting operators;
however none of these measurements have sharp expectation values. In
the present formulation we will find that the dispersion of any
coordinate introduces two distance regimes.
\begin{equation}\label{disp1}
\langle X^2\rangle -\langle X\rangle^2\sim\bar\theta + \frac{1}{\kappa^2}
\langle X\rangle^2\, ,
\end{equation}
with $\kappa$ a {\em very} large number and coordinate
noncommutativity determined by $\bar\theta$. For distances less than
$\kappa\sqrt{\bar\theta}$ the fluctuations in measurement of these
quantities are of order $\sqrt{\bar\theta}$ while for distances greater
than $\kappa\sqrt{\bar\theta}$ there is a fixed strain of $\sim
1/\kappa$. (The notation $\bar\theta$ and its relation to $\theta$
will be made apparent in the subsequent text.) As mentioned above,
$\kappa$ has to be very large -- how large will be discussed in
Section \ref{Conclusion}.

A general formulation of coordinate noncommutativity based on viewing
these coordinates as expectation values of operators between
generalized coherent states is presented in Section \ref{coord}. The
procedures for integration and for differentiation of functions of
coordinates, when these are treated as expectation values of operators
are discussed as is the implementation of translation invariance
through the introduction of a gauge potential. The subsequent three
sections are specific applications of Section \ref{coord} to various
dimensions and groups.

Noncommutativity in two spacial dimensions based on Heisenberg-Weyl
coherent states is discussed in Section \ref{H-W-Coh-States}. Many
modifications of quantum mechanics are the same as those obtained in
the star product formulation; it is the presence of fluctuations in
the measurement of any length and in the implementation of the
translation invariance by the introduction of a gauge potential
discussed above that makes the two approaches different. Although the
Heisenberg-Weyl group is not in the $SO(N,1)$ class this section
serves as a pedagogical introduction to the use of coherent states in
coordinate noncommutativity both in two and in higher number of
dimensions.

Two dimensional noncommutativity based on coherent states of the $2+1$
dimensional Lorentz group, $SO(2,1)\sim SU(1,1)$ is discussed in
Section \ref{SO(2,1)-Coh-States}. Many of the complications that will
occur when extending these ideas to higher dimensions will be
encountered here, but the algebra is sufficiently simple that results
are obtained in closed form. Section \ref{SO(3,1)-Coh-States} goes
through in detail the three dimensional case based on $SO(3,1)$ where
coordinate noncommutativity and rotational invariance coexist. In
Section \ref{Conclusion} extensions to dimensions greater than three
and the problems of time-space noncommutativity are noted. A
discussion of fluctuations in the measurement of any coordinate, as
presented in (\ref{disp1}), is given. Technical details of several
calculations may be found in the Appendixes.

\section{Coordinates as Coherent State Expectation
Values}\label{coord} Our goal is to obtain a theory invariant under
$N$ dimensional rotations. To this end we will consider coherent
states based on the groups, $SO(N,1)$ \cite{Perelomov}; an exception
will be for one of two versions of two dimensional noncommutative
quantum mechanics, the one based on the Heisenberg-Weyl algebra. We
should note that the Heisenberg-Weyl group is a contraction
\cite{Inonu:1953sp} of $SO(2,1)$. As taking over the general
discussions of this section to the Heisenberg-Weyl group are straight
forward the details will be presented for the $SO(N,1)$ cases. In
addition to the $SO(N)$ rotation operators $M_{ij}$, there are $N$
``noncompact'' operators $K_i$ whose commutation relations are
\begin{equation}\label{noncompcomm}
[K_i,K_j]=-iM_{ij}\, ;
\end{equation}
the minus sign in the above is crucial as it distinguishes this
algebra from compact $SO(N+1)$. We will be interested in unitary
representations of this algebra that contain all the representations
of $SO(N)$ starting with a one dimensional one; the states are labeled
as $|j,m;\kappa\rangle$, where $|j,m\rangle$ are irreducible
representations of $SO(N)$ and $\kappa^2$ is the value of the invariant
Casimir operator $K^2-\frac{1}{2}M_{ij}M_{ij}$; $j=j_0,
j_0+1,j_0+2,\ldots$ with the $j_0$ representation being one
dimensional; except for $N=2$, $j_0$ will be the zero angular momentum
state. The usual generalized coherent \cite{Perelomov} states are
defined as
\begin{equation}\label{cohst}
|\vec\eta; \kappa\rangle=e^{i\vec\eta\cdot\vec K}|j_0;\kappa\rangle\, .
\end{equation}

The next task is to determine the operators $\vec X$ which will
represent the coordinates. The measured value of $\vec X$ will be
$\langle\vec\eta; \kappa|\vec X|\vec\eta; \kappa\rangle$ and similarly
for any observable $F(\vec X)$ its measured value will be
$\langle\vec\eta; \kappa|F(\vec X)|\vec\eta;
\kappa\rangle$. This requirement puts several restrictions on the
choice of $\kappa$ and on the choice of the operators $\vec X$. 
\begin{itemize}
\item[(a)]
As $|\vec 0; \kappa\rangle$ will be taken to correspond to zero
length, $\langle\vec\eta; \kappa|\vec X|\vec\eta;
\kappa\rangle$ 
should be equal to $\hat\eta x(|\eta|)$ with $x(|\eta|)$ monotonic in
$|\eta|$ and $x(0)=0$. 
\item[(b)]
In order to control the fluctuations in the values of the coordinates
we require that 
\begin{equation}\label{fluct}
\langle\vec\eta; \kappa|X_i^2|\vec\eta;
\kappa\rangle - \left[\langle\vec\eta; \kappa|X_i|\vec\eta;
\kappa\rangle\right]^2 << \left[\langle\vec\eta;
\kappa|X_i|\vec\eta; \kappa\rangle\right]^2\, , 
\end{equation}
or more generally 
\item[(c)]
the cumulants \cite{Feller} of $\vec X$, defined as 
\begin{equation}\label{cumulant}
Q_{ijk\ldots}=\frac{-i\partial}{\partial
q_i}\ \frac{-i\partial}{\partial q_j }\ \frac{-i\partial}{\partial
q_k}\ldots \ln[\langle\vec\eta; \kappa|e^{i\vec q\cdot\vec
X}|\vec\eta;\kappa\rangle]|_{q_i=0}\, , 
\end{equation}
should satisfy $Q_{ijk\ldots}<< Q_{i}Q_{j}Q_{k}\ldots\/$. 
\end{itemize}
Conditions (b) and (c) require that the states we chose minimize
$\langle\vec\eta; \kappa|\vec X^2|\vec\eta; \kappa\rangle +
A\langle\vec\eta; \kappa|(\vec X\cdot\hat n)^2|\vec\eta;
\kappa\rangle$ subject to the constraint that $\langle\vec\eta;
\kappa|\vec X\cdot\hat n|\vec\eta;\kappa\rangle $ is fixed; $\hat
n$ is the direction along which we wish to measure $\vec X$ and the
constant $A$ in the above equation allows for a different expectation
value for $X^2$ along $\hat n$ and transverse to it. This naturally
leads to the variational problem of finding an eigenvector of the
operator $(\vec X)^2+A(\vec X\cdot\vec n)^2+\vec\lambda\cdot\vec X$,
with $\vec\lambda$ being a Lagrange multiplier fixing the average
value of $\vec X$. The coherent states, with $\vec\eta\sim\vec\lambda$
and $\kappa$ dependent on $A$, will be shown to be solutions of such
an eigenvalue problem; the choice of $\vec X$ will also lead to
condition (a) being satisfied by these states. Even though the
generalized coherent states are solutions of the above discussed
variational problem, we shall note that conditions (b) and (c) are
satisfied only for representations whose Casimir operators have {\em
very large values}. This condition also prevents large
momentum-momentum uncertainty relations. For two spacial dimensions,
$\vec X$ and $\vec K$ will be dual to each other.  {\em This will no
longer be the case for higher dimensions}\/.

A further complication arises for $N\ge 3$. It will be impossible to
satisfy conditions (b) and (c) above using expectation values of
position operators in coherent states as those discussed till
now. Taking a superposition of such matrix elements will solve this
problem. Namely we define
\begin{equation}\label{superposition}
<<\vec\eta|F(\vec X)|\vec\eta>> =
   \int d\kappa\, h(\kappa)\langle\vec\eta; \kappa|F(\vec X)|\vec\eta
    ;\kappa\rangle\, ,
\end{equation} for a suitably chosen $h(\kappa)$, with $\int d\kappa\,
h(\kappa)=1$; rather then taking expectation values in pure coherent
states $|\vec\eta;\kappa\rangle$ we do it in an ensemble described by
the density matrix $\rho=\int d\kappa h(\kappa)|\vec\eta;
\kappa\rangle\langle\vec\eta;\kappa|$. Expectation values defined in
this way {\em will} satisfy all the above conditions.

An extension of the concepts of integration over space and of
differentiation to the present situation in which coordinates are 
treated as expectation values of operators in coherent states is 
available. The over completeness \cite{Perelomov} 
of these states and the resolution of unity, $\int
\mu(\eta;\kappa)d\vec\eta\,  
|\vec\eta;\kappa\rangle\langle\vec\eta;\kappa|=1$, with
$\mu(\eta;\kappa)$ a 
representation dependent weight, permits the identifications 
\begin{eqnarray}\label{int-diff-ident}
\int d\vec x\,  f(\vec x)&\to& N_I\int
\mu(\eta;\kappa)d\vec\eta\, 
\langle\vec\eta; 
\kappa|f(\vec X)|\vec\eta; \kappa\rangle\, ,\nonumber\\   
\partial_jf(\vec x)&\to& N_D[iK_j, f(\vec X)]\, ,
\end{eqnarray}
with $N_I$ and $N_D$ constant; $N_I$ is chosen so that, for small
$\eta$, $N_I\mu(0;\kappa)\, d\vec\eta=d\vec x$ and $N_D$ is chosen to
yield $N_D\langle 0;\kappa|[iK_i,
X_j|0;\kappa\rangle=\delta_{ij}$. Some properties of this
``derivative'' have to be checked. The over completeness of the
coherent states makes $\int \mu(\eta;\kappa)d\vec\eta\,
\langle\vec\eta; \kappa|f(\vec X)|\vec\eta; \kappa\rangle$ proportional
to ${\rm Tr}\/ F(\vec X)$, where the trace is taken over a specific
representation. Thus we find that the integral of a derivative is
zero.  Likewise, the Leibniz rule is satisfied.

As mentioned earlier, we can extend invariance to the full $SO(N,1)$
group rather than just limiting it to the $SO(N)$ rotations generated
by the operators $M_{ij}$'s. Under the ``translations'' $\vec X
\to T^{\dag}(\vec a)\vec X T(\vec a)$, with $T(\vec a)=\exp
i\vec a\cdot\vec K$, completeness of the coherent states and the fact
that these are built on a one dimensional representation of the
rotation group guarantees
\begin{equation}\label{transinv1}
\int \mu(\eta;\kappa)d\vec\eta\, \langle\vec\eta;\kappa|T^{\dag}(\vec
a)f(\vec X)T(\vec a) 
|\vec\eta; \kappa\rangle\ = \int \mu(\eta;\kappa)d\vec\eta\,
\langle\vec\eta;  
\kappa|f(\vec X)|\vec\eta; \kappa\rangle\, .
\end{equation}
To maintain this invariance for expressions involving derivatives,
namely, $[K_j,f(\vec X)]$ a gauge potential ${\cal T}_j(\vec X)$ has
to be 
introduced with $[iK_j,f(\vec X)]$ replaced by $[iK_j-i{\cal T}_j(\vec
X),f(\vec X)]$  
and
\begin{equation}\label{transinv2}
{\cal T}_j(\vec X)\to T^{\dag}(\vec a){\cal T}_j(\vec X)T(\vec
a)-T^{\dag}(\vec a) [K_j,T(\vec a)]\, .
\end{equation}
The need to modify translations in the context of noncommutative
geometries has been discussed in \cite{Wess:2003da}. The field tensor
associated with ${\cal T}_j$ is
\begin{equation}\label{filedtensor}
{\cal F}_{ij}=[K_i,{\cal T}_j]-[K_j,{\cal T}_i]-[{\cal T}_i,{\cal T}_j]
  -[K_i,K_j]\, ;
\end{equation}
note that the last term has no analog for the case of ordinary
derivatives but is necessary for having ${\cal F}_{ij}$ transform
under rotations in the expected way, ${\cal F}_{ij}\to T^{\dag}(\vec
a){\cal F}_{ij}T(\vec a)$.

\section{Two Dimensional Noncommutativity Based on Coherent States of
the Heisenberg-Weyl Group}\label{H-W-Coh-States} 

The Heisenberg-Weyl algebra consists of the elements $K_i$, $i=1,2$
and $\mathbf 1$ with $[K_i,K_j]=i\epsilon_{ij}{\mathbf 1}$. The
coordinate operators are taken to be proportional to the dual of the
$K_i$'s 
\begin{equation}\label{coord1}
X_i={\sqrt\theta}\epsilon_{ij}K_j\, ,
\end{equation} 
resulting in the commutation relation
\begin{equation}\label{commut1}
[X_1,X_2]=i\theta\, .
\end{equation}

We look for a state in which the expectation of ${\vec X}$ is 
specified and the average of ${\vec X}\cdot {\vec X}$ is a minimum. The
standard variational principle leads us to look for an eigenstate of
${\vec X}\cdot{\vec X}-{\vec\lambda}\cdot {\vec X}$, where the
$\lambda_i$'s are Lagrange multipliers. With $|0\rangle$ annihilated
by $K_1+iK_2$, the coherent state 
\begin{equation}\label{coherst1}
|{\vec\eta}\rangle = e^{i{\vec\eta}\cdot{\vec K}}|0\rangle\, ,
\end{equation}
with $\vec\lambda=2{\sqrt\theta}\vec\eta$ is a solution of the
variational equation and 
\begin{equation}\label{expX1}
\langle\vec\eta|\vec X|\vec\eta\rangle ={\sqrt\theta}\vec\eta\, .
\end{equation}
From 
\begin{equation}\label{cummul1}
\langle\vec\eta|e^{i\vec q\cdot\vec X}|\vec\eta\rangle=
   e^{i{\sqrt\theta}\vec q\cdot\vec\eta-\frac{1}{4}\theta q^2}
\end{equation}
we can find the cumulants of $\vec X$, specifically
\begin{equation}\label{variation1}
\langle\vec\eta|X_iX_j|\vec\eta\rangle -
\langle\vec\eta|X_i|\vec\eta\rangle
\langle\vec\eta|X_j|\vec\eta\rangle =\delta_{ij}\frac{\theta}{2}
\end{equation}
and all higher cumulants are zero. As we wish to interpret the
expectation values of functions of $\vec X$ as measurements of these
quantities, it is gratifying that the fluctuations of these position
variables are under control. As mentioned in the introduction and in
Section \ref{coord}, this is a guiding principle in choosing states
and the operators $\vec X$.

Any classical function of the coordinates $f(x)=\int d{\vec
q}\, {\tilde f}(\vec q)\exp{i\vec q\cdot\vec x}$ may be transformed into
a similar integral with the c-numbers $\vec x$ replaced by the
operators $\vec X$. The product of exponentials of the $\vec X$'s
reproduces the star product rules in that
\begin{equation}\label{GMprod}
e^{i\vec k\cdot\vec X}e^{i\vec q\cdot\vec X}=
 e^{i(\vec k\cdot\vec X + \vec q\cdot\vec X)}
   e^{\frac{i\theta}{2}\epsilon_{ij}k_iq_j}\, .
\end{equation} 
Following (\ref{int-diff-ident}) we have a transcriptions of
integration over space and of differentiation 
\begin{eqnarray}\label{ident2}
\int d\vec x\, f(\vec x)&\to& \frac{\theta}{2\pi}\int d\vec\eta\, 
\langle\vec\eta|f(\vec X)|\eta\rangle\ , \nonumber\\
\partial_j f(\vec x)&\to& \frac{-i}{\sqrt\theta}
  [K_j, f(\vec X)]\, .
\end{eqnarray}
This also leads to the identification of momentum with the $K_i$'s
\begin{equation}\label{mom-ident}
\vec p\psi(\vec x)\to \frac{-1}{\sqrt\theta}[\vec K,\psi(\vec X)]
\end{equation}
which might lead one to worry that the $\sqrt\theta$ in the
denominator will introduce a large uncertainty relation for
momentum components. This is not the case \cite{Ambjorn:2000cs}
as we are making the
identification of momenta with commutators of $K$'s and by the Jacobi
identities
\begin{equation}\label{moment-uncert2}
[K_i,[K_j,\psi(\vec x)]]=[K_j,[K_i,\psi(\vec x)]\, .
\end{equation}
The observation that 
\begin{equation}\label{K-q2}
[-K_j,e^{i\vec q\cdot\vec X}]=\sqrt\theta q_ie^{i\vec q\cdot\vec X}
\end{equation}
as an operator identity and (\ref{GMprod}) leads to an algebra
identical to the star product one in that
\begin{eqnarray}\label{starequiv}
\frac{\theta}{2\pi}\int d\vec\eta\, \langle\vec\eta|[K_j,\psi(\vec X)]
[K_j,\psi(\vec X)]|\vec\eta\rangle & =& -\int d\vec x\, \partial_j\psi(x)
\partial_j\psi(x)\, ,\nonumber\\
\frac{\theta}{2\pi}\int d\vec\eta\, \langle\vec\eta|\psi_1(\vec X)
\ldots\psi_n(\vec X)|\vec\eta\rangle &=& \int d\vec x\, 
\psi_1(x)\star\ldots\star \psi_n(x)\, .
\end{eqnarray}

To see the restrictions of translation invariance it is useful to study
two particles interacting by a potential $V(\vec X^{(1)}-\vec X^{(2)})
=\int d\vec q\, \tilde V(\vec q)\exp(i\vec q\cdot \vec X^{(1)})
  \exp(-i\vec q\cdot \vec X^{(2)})$. The potential term in a
Lagrangian, 
\begin{equation}\label{Lv2}
{\cal L}_V=\frac{\theta^2}{4\pi^2}\int d\vec\eta^{(1)}\, 
 d\vec\eta^{(2)}\, \langle \vec\eta^{(1)},\vec\eta^{(2)}|
\psi^*(\vec X^{(1)}, \vec X^{(2)})V(\vec X^{(1)}-\vec X^{(2)})
\psi(\vec X^{(1)}, \vec X^{(2)})|\vec\eta^{(1)},\vec\eta^{(2)}
\rangle \, ,
\end{equation}
is invariant under the two particle analog of (\ref{transinv1})
\begin{equation}\label{transinv2a}
\psi(\vec X^{(1)}, \vec X^{(2)})\to T^\dag(\vec a)\psi(\vec X^{(1)},
\vec X^{(2)}) T(\vec a) \, ,
\end{equation}
with 
\begin{equation}\label{transinv2b}
T(\vec a)=e^{i\vec a\cdot\vec K^{(1)}}e^{i\vec a\cdot\vec K^{(2)}}\, .
\end{equation}
For the kinetic energy part to be invariant a gauge potential, ${\cal
T}(\vec X)$  transforming as in (\ref{transinv1}) has to be introduced. 
\begin{eqnarray}\label{Lk2}
{\cal L}_K&=&
\frac{-1}{4\pi^2}\int d\vec\eta^{(1)}\, 
 d\vec\eta^{(2)}\, \langle \vec\eta^{(1)},\vec\eta^{(2)}|\nonumber\\
&{}&\left\{
[K^{(1)}_j-{\cal T}_j(\vec X^{(1)}),\psi^*(\vec X^{(1)}, \vec
X^{(2)})][K^{(1)}_j-{\cal T}_j(\vec X^{(1)}),\psi(\vec X^{(1)}, \vec
X^{(2)})] \right. \\
&+& \left. [K^{(2)}_j-{\cal T}_j(\vec X^{(2)}),\psi^*(\vec X^{(1)}, \vec
X^{(2)})][K^{(2)}_j-{\cal T}_j(\vec X^{(2)}),\psi(\vec X^{(1)}, \vec
X^{(2)})]\right \}| \vec\eta^{(1)},\vec\eta^{(2)}\rangle\, ;
\nonumber 
\end{eqnarray}

The fact that the values of the coordinates are not sharp and the
inclusion of the translational gauge potential ${\cal T}(\vec X)$
differentiates this formalism from the star product one.

\section{Two Dimensional Noncommutativity Based on Coherent States 
of the $\bf{SO(2,1)\sim SU(1,1)}$ Group}\label{SO(2,1)-Coh-States}

Two dimensional coordinate noncommutativity can be obtained using
coherent states built on representations of the 2+1 dimensional
Lorentz group $SO(2,1)$ whose algebra is isomorphic to the one for the
$SU(1,1)$ group. Three generators $J,K_i$, with $i=1,2$, span this
algebra. $J$ is a rotation operator and the $K$'s are boosts. The
commutation relations are
\begin{eqnarray}\label{commrelSO(2,1)}
[J,K_i]& = &i\epsilon_{ij}K_j \nonumber\\
\left[K_i, K_j\right]& = &-i\epsilon_{ij}J\, ; 
\end{eqnarray}
the minus sign on the right hand side of the lower equation is
necessary to distinguish this algebra from the compact $SO(3)$
one. We shall be interested in the unitary, irreducible
representations of this group \cite{Bargmann} that consist of a tower
$|0;j\rangle, |1;j\rangle, |2; j\rangle,\ldots$ of one dimensional
representations of $J$ with with eigenvalues greater or equal to zero.
With $K_\pm=K_1\pm K_2$ the action of the operators is
\begin{eqnarray}\label{SO(2,1)ops}
J|n;j\rangle&=&(n+j)|j\rangle\nonumber\\
K_+|n;j\rangle&=&\sqrt{(n+1)(n+2j)}\, |n+1;j\rangle\\
K_-|n;j\rangle&=&\sqrt{n(n+2j-1)}\, |n-1;j\rangle\, .\nonumber
\end{eqnarray}
The Casimir invariant $J^2-\vec K\cdot\vec K=j(j-1)$. The coherent
states \cite{Perelomov,Barut} are
\begin{equation}\label{cohstate3}
|\vec\eta:j\rangle=e^{i\vec\eta\cdot\vec K}|0;j\rangle\, ,
\end{equation}
and the resolution of unity is
\begin{equation}\label{resolunit3}
1=\frac{2j-1}{4\pi}\int \sinh\eta\, d\eta\, d\hat\eta\,
|\vec\eta;j\rangle\langle\vec\eta;j|\, . 
\end{equation}

We now have to chose the operator that corresponds to the position
vector $\vec X$. The simplest choice is, as in the previous section,
$X_i=-\sqrt{\theta}\epsilon_{ij}K_j$; the minus sign is for subsequent
convenience. (In Appendix A we show that the coherent states are
solutions to the appropriate variational problem.) The expectation
value of $X_i$ can be obtained from the group algebra,
\begin{eqnarray}\label{measX3}
\langle\vec\eta;j|\vec X|\vec\eta;j\rangle &=&\langle 0;j|
\left[\vec X-{\hat\eta}{\hat\eta}\cdot\vec X + 
{\hat\eta}({\hat\eta}\cdot\vec X \cosh\eta + 
\sqrt{\theta}J\sinh\eta)\right]|0;j\rangle\nonumber\\
&=&j\sqrt{\theta}\hat\eta\sinh\eta\, .
\end{eqnarray}

It is immediately obvious that condition (b) in Section \ref{coord},
eq. (\ref{fluct}), is {\em not} satisfied as 
\begin{equation}\label{fluct3a}
\langle\vec\eta;j|(\hat\eta\cdot\vec X)^2 |\vec\eta;j\rangle -
\langle\vec\eta;j|\hat\eta\cdot\vec X |\vec\eta;j\rangle ^2=
\frac{j\theta}{2}(\cosh\eta)^2
\end{equation}
is of the same order as 
$\langle\vec\eta;j|\hat\eta\cdot\vec X |\vec\eta;j\rangle ^2$. More
generally 
\begin{equation}\label{cumulant3a}
\langle\vec\eta;j|e^{i\vec q\cdot\vec X}|\vec\eta;j\rangle=
\left(\cosh\frac{\sqrt{\theta}q}{2}-i\sinh\frac{\sqrt{\theta}q}{2}
\sinh\eta\, \hat q\cdot\hat\eta \right)^{-2j}\, ,
\end{equation} which does not lead to condition (c),
eq. (\ref{cumulant}). The desired properties can be recovered in the
large $j$ small $\theta$ limit, with $j\theta={\bar\theta}$ fixed. 
$\bar\theta$ sets the noncommutativity scale. To order $1/j$
(\ref{cumulant3a}) goes over to  
\begin{equation}\label{cumulant3b}
\langle\vec\eta;j|e^{i\vec q\cdot\vec X}|\vec\eta;j\rangle=
\exp\left[i\langle\vec\eta;j|\vec q\cdot\vec X|\vec\eta;j\rangle
-\frac{1}{4}\bar\theta q^2 - \frac{1}{4j}\langle\vec\eta;j|
\vec q\cdot\vec X|\vec\eta;j\rangle^2\right]\, . 
\end{equation}  
To lowest order in $1/j$ we reproduce (\ref{cummul1}) with $\theta$
replaced by $\bar\theta$ which, following the earlier discussion,
shows that in the large $j$ limit $SO(1,2)$ contracts to the
Heisenberg-Weyl group. To order $1/j$ (\ref{fluct3a}) takes on the form
\begin{equation}\label{fluct3b}
\langle\vec\eta;j|(\hat\eta\cdot\vec X)^2 |\vec\eta;j\rangle -
\langle\vec\eta;j|\hat\eta\cdot\vec X |\vec\eta;j\rangle ^2=
\frac{1}{2}\bar\theta +
\frac{1}{2j}\langle\vec\eta;j|\hat\eta\cdot\vec X |
\vec\eta;j\rangle^2 \, .
\end{equation}
This has the effect of introducing {\em two} very disparate distance
scales, $\sqrt{\bar\theta}$ and $\sqrt{j\bar\theta}$. For distances
less than $\sqrt{j\bar\theta}$ the fluctuations are fixed at the scale
of the noncommutativity parameter $\sqrt{\bar\theta}$, while for
distances on the order of or greater than $\sqrt{j\bar\theta}$ the
fluctuations grow as the distance itself divided by
$\sqrt{j}$. Clearly $j$ has to be very large. We will return to a
discussion of this point in Section \ref{Conclusion}.

The observation that $\langle\vec\eta;j|\vec X|\vec\eta;j\rangle =
j\sqrt{\theta}\sinh\eta\, \hat\eta$ and (\ref{int-diff-ident}) leads
to
\begin{eqnarray}\label{int-diff-ident3}
\int d\vec  x f(\vec x)&\to&j^2\theta \int\sinh\eta\, d\eta\, d\hat\eta 
\langle\vec\eta;j|f(\vec X) |\vec\eta;j\rangle\, ,\nonumber\\
\partial_jF(\vec x)&\to&\frac{-i}{j\sqrt{\theta}}[K_j,f(\vec X)]\, .
\end{eqnarray} 
Again, one has to check whether a large momentum-momentum uncertainty
has been introduced. Using Jacobi identities and the algebra of the
$K_i$'s and of $J$,
\begin{equation}\label{mom-mom3}
(\partial_i\partial_j-\partial_j\partial_i)\psi(\vec x)\to
\frac{-i}{j\bar\theta}[J,\psi(\vec X)]\, .
\end{equation}
As the expectation value of $[J,\psi(\vec X)]$ will be of the order
or less than that of $\psi(\vec X)$ the momentum-momentum uncertainty
will be of the order of $1/(j\bar\theta)$ which, by the previous
argument, will be very small.

\section{Three Dimensional Noncommutativity Based on Coherent States
of the $\bf{SO(3,1)}$ Group}\label{SO(3,1)-Coh-States}

As in the case of $SO(2,1)$ the generators three dimensional Lorentz
group, SO(3,1) break up into two classes, the compact ordinary angular
momenta $J_i$'s and noncompact $K_i$'s transforming as a vector
under angular momentum and with
\begin{equation}\label{K-comm4}
[K_i,K_j]=-i\epsilon_{ijk}J_k\, .
\end{equation}
The representations of $SO(3,1)$ \cite{Gelfand} are made up of towers
of the $(2j+1)$ dimensional unitary representations of $SO(3)$,
$|j,m\rangle$. We shall be interested in those that start with $j=0$
and thus contain
$|0,0;\kappa\rangle,|1,m:\kappa\rangle,|2,m;\kappa\rangle,\ldots$. The
value of the Casimir operator $\kappa^2=K^2-J^2$ (which we take to be
positive) determines the action of the operators $K_i$ on the angular
momentum states. The other quadratic Casimir operator $\vec K\cdot\vec
J=0$; in the notation of ref. \cite{Gelfand} we are dealing with
representations labeled by $l_0=0$ and $-l_1^2=\kappa^2+1$.

It is easy to note that choosing $\vec X$ to be linear in the
generators will not work. From the experience of the previous
sections, where a Levi-Civita symbol appeared in the definition of
$\vec X$, we are led to 
\begin{equation}\label{defX4}
X_i=\frac{\sqrt\theta}{2}\epsilon_{ijk}(J_jK_k+K_kJ_j)
\end{equation}
with 
\begin{equation}\label{3Dcommrel}
[X_i,X_j]=-i\theta\epsilon_{ijk}J_k(K^2+J^2)
\end{equation}
and 
\begin{equation}\label{3Ddiff-1}
[K_i,X_j]=-i\sqrt{\theta}\left[\delta_{ij}(K^2+J^2)-K_jK_i-J_jJ_i
\right]
\end{equation}
from which, as outlined in previous sections, we can define a
derivative. 

Let us first look at the expectation values in states of definite
$\kappa$ with $\kappa>>1$ \cite{strten}
\begin{eqnarray}\label{Xpar1}
\langle\vec\eta;\kappa|\hat\eta\cdot\vec X|\vec\eta;\kappa\rangle&=&
  \langle 0,0;\kappa|\cosh2\eta\, \hat\eta\cdot\vec X\nonumber\\
  &+&\frac{\sqrt\theta}{2}\sinh2\eta\left[\vec K^2-(\hat\eta\cdot\vec
    K)^2+\vec J^2-(\hat\eta\cdot\vec J)^2\right] 
    | 0,0;\kappa\rangle 
       =\frac{{\sqrt\theta}\kappa^2}{3}\sinh2\eta\, , \nonumber\\
    &{}&
\end{eqnarray}
while
\begin{eqnarray}\label{Xperp1}
\langle\vec\eta;\kappa|\hat\eta\times\vec X|\vec\eta;\kappa\rangle&=&
\langle 0,0;\kappa|\cosh\eta\,\hat\eta\times\vec X\nonumber\\
&-&\frac{\sinh\eta}{2}
\left(\hat\eta\times\vec K\hat\eta\cdot\vec K-
\hat\eta\cdot\vec J\hat\eta\times\vec J\right)|0,0;\kappa\rangle
=0\, .
\end{eqnarray}
Details of the proof that these coherent states minimize $\vec X^2+
A(\hat\eta\cdot\vec X)^2$, with $\vec X$ defined in (\ref{defX4}), are
given in Appendix B.

The leading, in $\kappa^2$, contributions to the expectation values of
$(\hat\eta\cdot\vec X)^n$ will come from $\langle 0,0;\kappa|( \vec
K^2-(\hat\eta\cdot\vec K)^2)^n|0,0;\kappa\rangle $. In order to satisfy
conditions (b) an (c) of Section \ref{coord} this matrix element has
to be equal to $(2\kappa^2/3)^n$. Such a condition was satisfied for
the two dimensional case discussed in Section \ref{SO(2,1)-Coh-States}
but due to the $(\hat\eta\cdot\vec K)^2$ terms it is not satisfied in the
present three dimensional formulation. These matrix elements are
evaluated in the Appendix C, (\ref{appres}). To leading order in $\kappa$
\begin{equation}\label{kappadisp1}
\langle\vec\eta;\kappa|(\hat\eta\cdot\vec X)^n|\vec\eta;\kappa\rangle=
 \frac{\sqrt\pi}{2}\frac{\Gamma(n+1)}{\Gamma(n+\frac{3}{2})} 
  \left[\frac{3}{2}\langle\vec\eta;\kappa|\hat\eta\cdot\vec X
          |\vec\eta;\kappa\rangle\right]^n\, .
\end{equation}
This is unacceptable as it would lead to large fluctuations in the
expectation values of the position operator. With the observation
\begin{equation}\label{observ}
\kappa_0^{2n}=\frac{\sqrt\pi}{2}\frac{\Gamma(n+1)}{\Gamma(n+\frac{3}{2})}
\frac{1}{\pi}\int_0^{\kappa_0}\frac{d\kappa}{\sqrt{\kappa_0^2-\kappa^2}}
\frac{\partial}{\partial\kappa}\kappa^{2n+1}
\end{equation}
we follow the discussion outlined around (\ref{superposition}) and
consider a new averaging procedure defined by double bras and kets
\begin{equation}\label{newexp1}
<<\vec\eta|{\cal O}(\vec J,\vec K)|\vec\eta>>=
\frac{1}{\pi}\int_0^{\kappa_0}\frac{d\kappa}{\sqrt{\kappa_0^2-\kappa^2}}
\frac{\partial}{\partial\kappa}\kappa
 \langle\vec\eta;\kappa|{\cal O}(\vec J,\vec K)|\vec\eta;\kappa\rangle.
\end{equation}
In (\ref{observ}) and (\ref{newexp1}) $\kappa_0>>1$ replaces $\kappa$
as a parameter determining the representations of $SO(3,1)$ used.
(The lower limit in these integration can be replaced by $\kappa_1$ as
long as $\kappa_1/\kappa_0<<1$.) 

With this new choice of states we have
\begin{eqnarray}\label{newexptval}
<<\vec\eta|\hat\eta\cdot\vec X|\vec\eta>>&=&
  \frac{\sqrt{\theta}\kappa_0^2}{2}\sinh 2\eta\, ,\nonumber\\
<<\vec\eta|(\hat\eta\cdot\vec X)^2|\vec\eta>>-
 <<\vec\eta|\hat\eta\cdot\vec X|\vec\eta>>^2 &=&
  \theta\kappa_0^2+ \frac{4}{\kappa_0^2}
   <<\vec\eta|\hat\eta\cdot\vec X|\vec\eta>>^2\, ;
\end{eqnarray}
As in Section \ref{SO(2,1)-Coh-States} we go to the large $\kappa_0$ 
limit with $\theta\kappa_0^2=\bar\theta$ fixed and it is
$\bar\theta$ that sets the coordinate noncommutativity scale.
To order $1/\kappa^2$ 
\begin{equation}\label{fluct4} 
<<\vec\eta;j|(\hat\eta\cdot\vec X)^2 |\vec\eta;j>> -
<<\vec\eta;j|\hat\eta\cdot\vec X |\vec\eta;j>> ^2=
\bar\theta +
\frac{4}{\kappa_0^2}<<\vec\eta;j|\hat\eta\cdot\vec X |
\vec\eta;j>>^2 \, .
\end{equation}
Again, there are two distance scales.  For distances less than
$\kappa_0\sqrt{\bar\theta}$ the fluctuations are fixed at the
noncommutativity parameter $\sqrt{\bar\theta}$, while for distances on
the order of or greater than $\kappa_0\sqrt{\bar\theta}$ the
fluctuations grow as the distance itself divided by
$\kappa_0$. Discussion of these results is left for the next section.

\section{Conclusion}\label{Conclusion} We have formulated a coordinate
noncommuting quantum mechanics where the measurement of position
operators, or functions of such operators, is determined by their
expectation values between generalized coherent states. The concepts
of integration and of differentiation can be incorporated. This
formulation is invariant under the full rotation group and translation
invariance holds at the price of introducing a gauge potential. As the
measurement of none of the coordinate components is sharp, special
attention has to be paid to control any of the fluctuations in the
values of these components. This is achieved only for the coherent
states built on representations with very large values of the relevant
Casimir operators.

Details were presented for two and three dimensions. Extensions to
higher dimensions, though algebraically tedious, are straight forward.
Coherent states for $SO(N,1)$ groups with $N>3$ are discussed in
\cite{Perelomov}. The position operators may be taken as
\begin{equation}\label{NdimX}
X_i=\frac{\sqrt{\theta}}{2}(M_{ij}K_j+K_jM_{ij})\, ,
\end{equation}
where the $M_{ij}$'s are the generators of the $SO(N)$ subgroup of
$SO(N,1)$. Extension to where time is one of the noncommuting directions is problematical not only for general reasons having to do with
violations of unitarity \cite{Unit-viol} but also due to technical
difficulties of extending the present formalism. Naively, to introduce
time as one of the noncommuting coordinates one might try to construct
coherent states based on the de Sitter group, $SO(N,2)$ with $SO(N,1)$
being the symmetry group and the space-time operators in the coset
$SO(N,2)/SO(N,1)$. (It is amusing to note that one of the early
attempts at noncommutativity \cite{Snyder} placed the space-time
coordinates into such a coset space.) Although such coherent states
have, to this authors knowledge, not been studied one problem can be
seen immediately: time becomes periodic. This may be noted easily by
looking at the previously studied group $SO(1,2)$, now thought of as a
de Sitter group.

An intriguing result of this work is displayed in (\ref{disp1}) where
fluctuations in the measurement of any coordinate introduce two
distance scales. The first scale is the expected one set by the
noncommutativity parameter $\sqrt{\bar\theta}$ while the other one
depends on the value of the Casimir operator $\kappa^2$ and is equal
to $\sqrt{\bar\theta}\kappa$ and induces fluctuations proportional to
the length itself. This relation, summarized in (\ref{disp1}) is
reminiscent of the generalized Heisenberg uncertainty relation
\cite{genuncert}
\begin{equation}\label{genunc}
\Delta x\Delta p\ge \frac{\hbar}{2}+
    \frac{\theta (\Delta p)^2}{\hbar}\, .
\end{equation} $\sqrt{\bar\theta}$ is presumably very small, possibly
of the order of the Planck length $\lambda_P$; $\kappa$ which acts
like a strain will have to be large. Limits of $\kappa > 10^{21}$ will
be available from measurements at the LIGO Observatory
\cite{Barish:1999vh}.

\appendix
\section{Coherent States as Solutions of The $SO(2,1)$ Variational
Problem}
With the choice for coordinate operators made in Section
\ref{SO(2,1)-Coh-States} we will show that the coherent states
minimize the the expectation of $\vec X^2+A(\hat\eta\cdot\vec X)^2$
with the expectation of $\vec X$ fixed. We may take $\hat\eta$
along the $x$ direction. We , thus, have to show that for some $\eta$
and $j$ $|\eta\hat x; j\rangle$ satisfies the eigenvalue
equation
\begin{equation}\label{eigen-II.1}
[\vec X^2 +A(\hat x\cdot\vec X)^2+\lambda\hat x\cdot X]|\eta\hat x;
j\rangle = c|\eta\hat x;j\rangle
\end{equation}
with $\lambda$ a Lagrange multiplier and $c$ the eigenvalue. This
is equivalent to showing 
\begin{equation}\label{eigen-II.2} 
  e^{-i\eta K_x}[\vec X^2 +A(\hat x\cdot\vec X)^2-
\lambda\hat x\cdot X] e^{i\eta K_x}|0;j\rangle =c|0;j\rangle.
\end{equation}
The unitary transformation on the left hand side of the above can be
carried out explicitly and using the fact that 
$K_-|0;j\rangle=0$ we have to set the coefficients of $K_+^2$ and
$K_+$ to zero. This leads to the conditions
\begin{eqnarray}\label{condA.1}
\frac{1}{4}\left[(1+A)\cosh^2\eta-1\right]K_+^2& = &0\, ,\nonumber\\
(2j\sinh\eta+\lambda)K_+ & = &0\, ,
\end{eqnarray}
or, with $\langle\hat x\cdot\vec X\rangle=j\sqrt{\theta}\sinh\eta$,
we find
\begin{equation}\label{condA.2}
A=-\frac{\langle\hat x\cdot\vec X\rangle}
  {\sqrt{\langle\hat x\cdot\vec X\rangle^2+j\bar\theta}}\, .
\end{equation}

\section{Coherent States as Solutions of The $SO(3,1)$ Variational
Problem}
We will follow the same procedure for $N=3$ as we did in Appendix A
for $N=2$; this time we take $\vec\eta$ to be along the $\hat z$
direction and show that there are values of $\eta$ and of
$\kappa$ that lead to a solution of the eigenvalue problem 
\begin{equation}\label{eigen-III.1}
e^{-i\eta K_z}\left[\vec X^2+A (\hat z\cdot\vec X)^2+\lambda
  \hat z\cdot\vec X\right]e^{i\eta K_z}
|0,0;\kappa\rangle= c|0,0;\kappa\rangle \, .
\end{equation}
After performing the unitary transformation and noting that
$|0,0;\kappa\rangle$ is annihilated by each component of $\vec J$ we
set the coefficients of $K_z^4$ and of $K_z^2$ to zero, 
\begin{eqnarray}\label{cond-III.1}
\sinh^2\eta\left[(1+A)\cosh^2\eta-1\right]K_z^4&=&0\, ,\nonumber\\
\left[\left(\cosh^22\eta-\frac{\kappa^2}{2}\sinh^22\eta\right)-
  \lambda\sinh2\eta\right]K_z^2&=&0\, .
\end{eqnarray}
Again, with $\langle\hat z\cdot\vec X\rangle
=(\sqrt{\theta}\kappa^2\sinh2\eta)/3$, $A$ is related to $\kappa$,
\begin{equation}\label{cond-III.2}
A=\left[{1-\sqrt{1+\frac{9\langle\hat z\cdot\vec X\rangle^2}
   {\kappa^2\bar\theta}}}\right]
   \left[{1+\sqrt{1+\frac{9\langle\hat z\cdot\vec X\rangle^2}
   {\kappa^2\bar\theta}}}\right]^{-1}
\end{equation}

\section{Evaluation of Certain $SO(3,1)$ Matrix Elements}
The matrix element $\langle 0,0;\kappa|\left[\vec K^2-(\hat\eta\cdot\vec
K)^2\right]^n|0,0;\kappa\rangle$, for $n<\kappa$, needed in Section
\ref{SO(3,1)-Coh-States} will be evaluated. By rotational invariance
we may set $\hat\eta=\hat z$ and evaluate $\langle 0,0;\kappa|(\vec
K^2-K_z^2)^n|0,0;\kappa\rangle$. The states $|l,0;\kappa\rangle$ are
eigenstates of $\vec K^2$ with eigenvalue $\kappa^2
+l(l+1)\sim\kappa^2$. In order to obtain $\langle
0,0;\kappa|K_z^{2\mu}|0,0;\kappa\rangle$, with $\mu$ integer we
evaluate the coefficients $\alpha_l^{(2\mu)}$ in the expansion
\begin{equation}\label{appexp}
K_z^{\mu}|0,0;\kappa\rangle=\kappa^{\mu}\sum_l(-i)^l\sqrt{2l+1}
\alpha_l^{(\mu)}|l,0;\kappa\rangle\, ;
\end{equation}
$l$ is restricted to even values. For the representation of interest
and for $\kappa>>l$ the action of $K_z$ on the states
$|l,0;\kappa\rangle$ 
is \cite{Gelfand}
\begin{equation}\label{actKz}
K_z|l,0;\kappa\rangle=i\kappa\left(\frac{l}{\sqrt{4l^2-1}}|l-1,0;
\kappa\rangle-\frac{l+1}{\sqrt{4(l+1)^2+1}}
|l+1,0;\kappa\rangle\right)\, .
\end{equation}
It is straightforward to obtain the recursion relation for the
$\alpha_l^{(\mu)}$'s ($\alpha_0^{(0)}=1$),
\begin{equation}\label{recrel}
(2l+1)\alpha_l^{(\mu+1)}=(l+1)\alpha_{l+1}^{(\mu)}+
l\alpha_{l-1}^{(\mu)}\, ,
\end{equation}
whose solution is 
\begin{equation}\label{solnrec}
\alpha_l^{(\mu)}=\frac{1}{2}\frac{(\frac{\mu}{2})!(\frac{\mu-1}{2})!}
{(\frac{\mu-l}{2})!(\frac{\mu+l+1}{2})!}
\end{equation}
and especially
\begin{equation}\label{alpha0}
\alpha_0^{(\mu)}=\frac{1}{\mu+1}\, .
\end{equation}
From the above we obtain
\begin{eqnarray}\label{appres}
\langle 0,0;\kappa|(\vec K^2-K_z^2)^n|0,0;\kappa\rangle&=&
\kappa^{2n}\sum_{\mu}(-1)^\mu\frac{n!}{\mu !(n-\mu)!}\frac{1}{2\mu+1}
\nonumber\\
&=&\kappa^{2n}\frac{\sqrt\pi}{2}\frac{\Gamma(n+1)}{\Gamma(n+\frac{3}{2})}
\end{eqnarray}


\begin{thebibliography}{99}
\bibitem{strten}
One way these transformations can be understood is to note that there
is an analogy between $\vec K, \vec J$ and $\vec E, \vec H$ of
electromagnetism. $\vec X$ is the analog of the Poynting vector or the
$T_{0i}$ components of the energy-momentum stress tensor; $T_{00}\sim
(\vec J^2+\vec K^2)/4$ and $T_{ij}\sim \delta_{ij}(\vec J^2+\vec
K^2)/4 -(J_iJ_j+K_iK_j)/2$.
\end{thebibliography}
\end{document}